\documentstyle[twocolumn,eclepsf,epsfig,graphics]{jpsj}

\def\runtitle{3K-phase in Sr$_2$RuO$_4$}
\def\runauthor{Manfred {\sc Sigrist} and Hartmut {\sc
Monien}}

\title{Phenomenological theory of the 3 Kelvin phase in Sr$_2$RuO$_4$}  

\author{Manfred {\sc Sigrist} and Hartmut {\sc Monien}\footnote{
Permanent address: Physikalisches Institut, Universit\"at Bonn,
Nussallee 12, 53115 Bonn, Germany}}
\inst{Yukawa Institute for Theoretical Physics, Kyoto University,
Kyoto 606-8502, Japan}

\recdate
{\today}

\abst{We model the 3K-phase of Sr$_2$RuO$_4$ with Ru-metal inclusion as
interface state with locally enhanced transition temperatures. The resulting
3K-phase must have a different
pairing symmetry than the bulk phase of Sr$_2$RuO$_4$, because the
symmetry at the interface is lower than in the bulk. It is
invariant under time reversal and a second transition, in general, above the onset
of bulk superconductivity is expected where time reversal symmetry
is broken. The nucleation of the 3K-phase exhibits a
``capillary effect'' which can lead to frustration phenomena for the
superconducting states on different Ru-inclusions. Furthermore, the
phase structure of the pair wave function gives rise to zero-energy
quasiparticle states which would be visible in quasiparticle tunneling spectra. 
Additional characteristic properties are associated with the upper
critical field $ H_{c2} $. The 3K-phase has a weaker anisotropy of $
H_{c2} $ between the inplane and $z$-axis orientation than the bulk
superconducting phase. This is connected with the more
isotropic nature Ru-metal which yields a stronger orbital depairing effect
for the inplane magnetic field than in the strongly layered
Sr$_2$RuO$_4$. An anomalous temperature
dependence for the $z$-axis critical field is found due to the
coupling of the magnetic field to the order parameter texture at the
interface. Various other experiments are discussed and new measurements are suggested.}

\kword{Sr$_2$RuO$_4$, $p$-wave superconductor, broken time-reversal
symmetry, frustration effects, Andreev bound states}

\begin{document}
\sloppy
\maketitle

\section{Introduction}

In recent years the quasi-two-dimensional metal Sr$_2$RuO$_4$ has advanced to
one of the most intensely studied transition metal oxides displaying
unconventional superconductivity.\cite{MAENO1,NATRICE} There is strong
evidence for spin-triplet pairing with broken time reversal symmetry,
a pairing state with the basic form $ {\bf d} ({\bf k}) =
\hat{{\bf z}} (k_x \pm i k_y) $ (chiral p-wave
state).\cite{RICE,ISHIDA1,LUKE} This superconducting phase  
shows a variety of unusual properties among which the recently discovered 
``3-Kelvin'' (3K) phase is one of the most puzzling findings. While
the transition to the bulk superconducting state occurs around $ 1.5 $K, 
in samples with a large excess-Ru concentration a
precursors to the superconducting transition appears at temperatures
as high as 3K.\cite{3K,ANDO} This phase shows the features of an inhomogeneous
superconducting phase. The detailed material analysis revealed that
the excess-Ru does not distribute uniformly, but forms small inclusions of 
micrometer-size. Thus, the 3K-phase is likely connected with the
phase separation of Ru-metal and Sr$_2$RuO$_4$ in this sample. The bulk
superconductivity of metallic Ru has a superconducting transition
temperature of 0.5 K only with a conventional ($s$-wave)
pairing state. This leads us to the assumption that the boundaries between the
two materials provides the environment for the local nucleation of
superconductivity at a higher temperature (Fig.~1).

It is not our aim to discuss here in detail the origin for the locally
enhanced transition temperature, since the microscopic theory of
superconductivity in Sr$_2$RuO$_4$ is still unclear. Nevertheless, we
would like to comment on one important aspect which could be connected
with the enhanced superconductivity. Various experiments have shown
that a particular soft optical modes for local lattice distortion is
associated with the inplane rotation of the RuO$_6$-octahedra, the $
\Sigma_3 $-mode at the Brillouin zone boundary of the phonon spectrum,
which leads to a slight volume reduction.\cite{NEUTRON} This rotation
affects one of the three electron bands in particular, the $ \gamma
$-band which originates from the $4d$-$t_{2g}$-orbital with $ d_{xy}
$-symmetry.\cite{RICE,LDA} While the dispersion for this orbital
occurs via $ \pi $-hybridization between the Ru-$d$-orbital and the
O-$p$-orbitals in the undistorted case, the rotation introduces an
additional $ \sigma $-hybridization of opposite sign for the $d_{xy}
$-orbital. A lattice distortion of this kind would diminish the
dispersion of the $ \gamma $-band and enhance the electron density of
states, because of a Van Hove singularity near the Fermi level.
\cite{TERAKURA}.  It is likely that the Ru-inclusions in Sr$_2$RuO$_4$
lead to internal stress that is released by local static distortions
in the vicinity of the interfaces, most likely connected with the $
\Sigma_3$-mode ($ \hbar \omega_{\Sigma_3} \approx 140 $K). A crude
estimate from neutron scattering data yields a length scale of order 
50 - 100\AA$ $ over which this rotational distortion extend away
from the interface. The increased density of states would lead to a
locally enhanced $ T_c$ independent of the microscopic mechanism. The
additionally enhanced ferromagnetic spin fluctuations may support the
spin-triplet pairing instability too.\cite{TERAKURA}

In this paper we would like to investigate a number of properties of
the inhomogeneous 3K-phase from a phenomenological point of view.  A
generalized Ginzburg-Landau (GL) formulation is most suitable for this
purpose, since we will discuss an inhomogeneous superconducting phase.
The basic assumption is that the interface region has an enhanced
transition temperature. We briefly review the basic conclusions of our
theory. The first and most important fact is that the superconducting
state nucleated at the interface has a different symmetry than the
bulk superconducting phase. The interface superconducting state is
time reversal symmetry conserving. Consequently, there is a further
second order phase transition where this symmetry is broken. We will
show that this transition occurs in general above the onset of bulk
superconductivity. Naturally the nucleation of superconductivity at
the interface is inhomogeneous and does not lead to a uniform phase
transition. This is also true for the second transition. Even if all
interfaces between Ru-inclusions and Sr$_2$RuO$_4$ are locally
equivalent, their geometry and mutual arrangement would lead to a
spread on nucleation temperatures due to ``capillary effects''.

The 3K superconducting state at the interface corresponds to an odd
parity state, a $p$-wave state with a pair wave function that has a
node parallel to the normal vector with a positive and negative lobes
parallel to the interface (Fig.~1). This phase structure is responsible for 
a peculiar change of the quasiparticle spectrum due to Andreev
reflection, i.e. the accumulation of Andreev bound states at
zero-energy which may be observed by quasiparticle tunneling. 
The phase structure together with the topology of the interfaces 
can generate a frustration of the order parameter phase which can be
released by introducing spontaneous orbital currents. Note that also
the second transition to a time reversal symmetry breaking phase would
reduce the frustration and introduce spontaneous currents. 

Finally the upper critical field $ H_{c2} $ shows special properties
in case of fields parallel to the interface. In this case the orbital
depairing is reduced due to the essentially two dimensional nature of
the condensate. In addition, the anisotropy of $ H_{c2} $ between
fields parallel to basal plane of the strongly layered Sr$_2$RuO$_4$
and the $z$-axis direction is reduced compared with the critical field
for the bulk phase. This is due to the more isotropic nature of the
Ru-inclusions. A considerable fraction of the superconducting
condensate of the 3K-phase resides in the Ru-part. An additional very
striking feature occurs for magnetic field parallel to the $z$-axis of
Sr$_2$RuO$_4$. While in zero field only the $p$-wave component with
its node perpendicular to interface appears, the magnetic field drives
also the other component whose node is parallel to the interface. The
energy gain occurs via the coupling of the field to the current
induced by the order parameter texture of both components. This
coupling leads to a peculiar enhancement and temperature dependence of
$ H_{c2} $ in agreement with experiment.\cite{3K,ANDO}

\section{Phenomenological description}

The following analysis is based on the generalized GL theory as the
most efficient way to describe basic properties of an inhomogeneous
superconducting state. The bulk pairing state of Sr$_2$RuO$_4$ has the
symmetry of a chiral $ p $-wave state, represented by $ {\bf d} ({\bf
  k}) = \Delta_0 \hat{{\bf z}} (k_x \pm i k_y) $. This requires a
two-component order parameter $ \mbox{\boldmath $ \eta $} = ( \eta_x ,
\eta_y ) $ with $ {\bf d} ({\bf k}) = \hat{{\bf z}} (\mbox{\boldmath $
  \eta $} \cdot {\bf k}) $, which belongs to the two-dimensional
representation $ E_u $ of the tetragonal point group $ D_{4h} $ and
describes the leading instability in Sr$_2$RuO$_4$. Note that recent
flux distribution measurements in the mixed state have suggested the
presence of two order parameter components.\cite{fluxdist} On the
other hand, Ru is a conventional s-wave superconductor with a
transition temperature around 0.5 K.  The superconductivity in
Sr$_2$RuO$_4$ penetrates the Ru-metal due to the proximity effect in
two ways.  First, proximity leads naturally to a spin-triplet pairing
amplitude in Ru, although its critical temperature there may be
extremely small.  Second, the spin-triplet superconducting state in
Sr$_2$RuO$_4$ can induce the s-wave component. The corresponding
coupling is, however, probably weaker than for the triplet channel
because at the interface the triplet and singlet spin wave function
have to be connected, by means of spin-orbit scattering.  We will
ignore the s-wave component in Ru and will briefly comment later only.

\subsection{Ginzburg-Landau free energy}

The GL free energy for
the two-component order parameter $ \mbox{\boldmath $ \eta $} $ has
the well-known form, 
\begin{equation}
\begin{array}{ll}
{\cal F} =& \displaystyle \int d^3 r [ a | \mbox{\boldmath $ \eta
$}|^2 + b_{1}  
| \mbox{\boldmath $ \eta $}|^4 + \frac{b_2}{2}(\eta^{*2}_x \eta_y^2 +
c.c.) \\ &  \displaystyle + b_3 |\eta_x|^2 |\eta_y|^2  
+ K_{1} (|D_x \eta_x|^2 +  |D_y
\eta_y|^2) \\ & \displaystyle  + K_{2} (|D_y \eta_x|^2 
+ |D_x \eta_y|^2)
+ \{K_{3} (D_x \eta_x)^*(D_y \eta_y)  \\ &  + K_{4}
((D_y \eta_x)^*(D_x \eta_y) + c.c.\} \\ & \displaystyle 
+ K_{5} (|D_z \eta_x|^2 + | D_z \eta_y|^2) +
(\nabla \times {\bf A})^2/8 \pi ] 
\end{array} \label{free-energy}\end{equation}
where the coefficients are different in the two subsystems which we
label by indices S and R for Sr$_2$RuO$_4$ and Ru, respectively.  
The gradient terms contain the gauge invariant spatial
derivatives $ {\bf D} = \nabla + i (2e/\hbar c) {\bf A} $ with $ {\bf
A} $ denoting the vector potential. 
The coefficients $ K_{i \mu} $ determine the coherence length of the
superconducting order parameter ($ \mu = $ S and R). Since
Sr$_2$RuO$_4$ has a layered structure, the coherence length along the
z-axis is short, $ K_{5 S} \ll K_{1 S}, K_{2 S}, ... $. On the
other hand, Ru is more isotropic so that $ K_{5R} $ has a magnitude
similar to the other coefficients.  
In a weak-coupling approach assuming cylindrical or spherical Fermi
surface shapes we obtain the relation $ K_{1 \mu} / 3 = K_{2 \mu} =
K_{3 \mu} = K_{4 \mu} $. 
The second order coefficient  $ a_\mu (T) $ changes sign at the bare
bulk transition temperature $ T_{c \mu} $. For our discussion of the
qualitative properties of the interface superconductivity it is
sufficient to assume linear temperature dependence $ a_\mu (T) =
\alpha_\mu (T - T_{c \mu}) $ with $ T_{c {\rm S}} \approx 1.5 $K and $
T_{c {\rm R}} = 0 $.  

For simplicity, we consider a single homogeneous planar interface with
a normal vector parallel to the $ x $-axis ($ {\bf n} = (100) $) where
$ -\infty < x <0 $ belongs to the Ru-metal, and $ 0 < x < \infty $ to
Sr$_2$RuO$_4$. In this geometry the problem reduces to a one
dimensional with the spatial direction along the $x$-axis.
In the following we will always ignore the $z$-direction assuming
homogeneity in this direction for simplicity. On the
Sr$_2$RuO$_4$-side a thin layer of thickness $d$ with enhanced
transition temperature is introduced. The boundary conditions, in
general, involve reflection and transmission of Cooper pairs at the
interface. We make the simplifying assumption of complete
transparency, which corresponds to a continuous order parameter. We
will, however, comment on the more general case below.  We may
describe the thin layer at the interface by a $ \delta $-function in
the free energy, if $ d $ is much smaller than the coherence lengths
along the $ x $-axis (Fig.~1). Thus, the interface part of the free
energy has the form,
\begin{equation}
{\cal F}_i = \int d^3 r \delta(x) \sigma | \mbox{\boldmath $ \eta
$}|^2
\end{equation}
where $ \sigma = d \alpha_{\sigma} (T-T_{c \sigma}) $ with $ T_{c \sigma} > 
T_{c {\rm S}} $. This approximation is sufficient to discuss most
important qualitative features of the 3K-phase. 

\begin{figure}
\begin{center}
\includegraphics*[height=5.0cm,width=6cm]{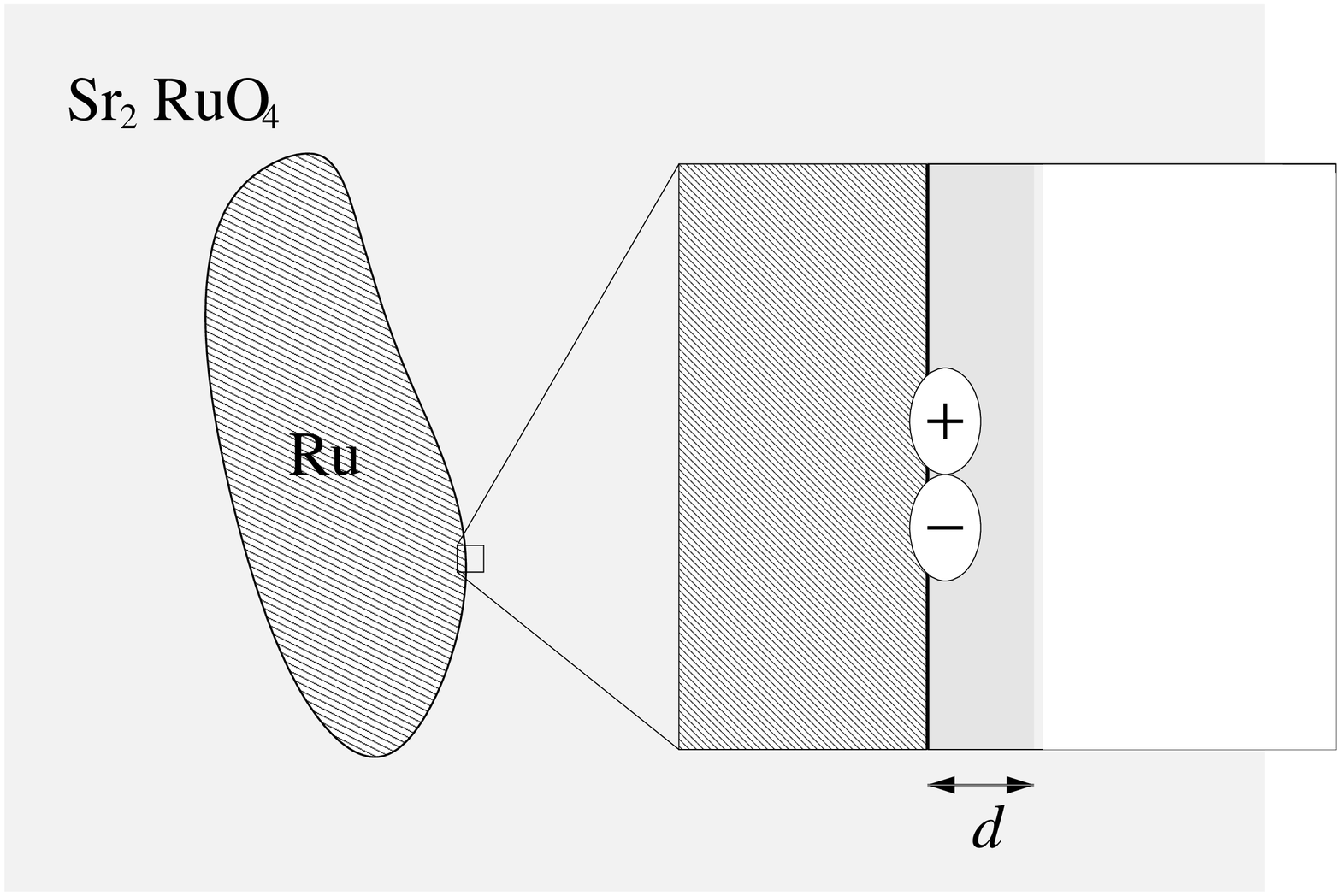}
\end{center}
\caption{Interface between Sr$_2$RuO$_4$ and a Ru-inclusion. The
interface has a layer of thickness $ d$ of enhanced transition
temperature where a $ p$-wave state nucleates whose wavefunction 
has the lobes parallel and the nodes perpendicular to
the interface.}
\end{figure}

\subsection{Instability conditions of a planar interface}

We will first investigate the conditions of
the nucleation of superconductivity at the planar interface. We consider
a temperature range $ T_{c {\rm S}} < T < T_{c \sigma} $ where this
local onset of superconductivity is supposed to occur. The instability 
condition is obtained by searching for the solution of the linearized
GL equations which are given by
\begin{eqnarray} 
K_{1\mu} \partial_x^2 \eta_x - a_{\mu} \eta_x = 0,  \\
K_{2\mu} \partial_x^2 \eta_y - a_{\mu} \eta_y = 0,
\end{eqnarray}
for $ x \neq 0 $ ($ \mu = $ S and R for $ x >0 $ and $ x < 0 $,
respectively). At $ x=0 $ the solutions have to be continuous and 
satisfy the following boundary conditions,
\begin{eqnarray}
K_{1{\rm S}} \partial_x \eta_x|_{x=0_+} - K_{1{\rm R}} \partial_x
\eta_x |_{x=0_-} - \sigma \eta_x|_{x=0} = 0 ,
\label{bc1} \\
K_{2{\rm S}} \partial_x \eta_y|_{x=0_+} - K_{2{\rm R}} \partial_x
\eta_y |_{x=0_-} - \sigma \eta_y|_{x=0} = 0.
\label{bc2}
\end{eqnarray}
For temperatures above $ T_{c {\rm S}} $ the order parameter is
largest at the interface and decays exponentially on both sides. For the 
$ \eta_x $-component the solution is
\begin{eqnarray}
\eta_x (x) =  \eta_{x0} e^{-x / \xi_{1 {\rm S}}} & \mbox{  for  } x > 0 
\label{exps}
\\
\eta_x (x) =  \eta_{x0} e^{x / \xi_{1 {\rm R}}} & \mbox{  for  } x < 0 
\label{expr}
\end{eqnarray}
with $ \xi_{1 \mu}^2 = K_{1 \mu} / a_{\mu} $. The analogous solution
exists for $
\eta_y $ with $ \xi_{2 \mu }^2 = K_{2 \mu}/a_{\mu} $. Then
Eq.(\ref{bc1}) and (\ref{bc2}) lead to the instability equations,
\begin{eqnarray}
\sqrt{K_{1 {\rm S}} a_{\rm S}} + \sqrt{K_{1 {\rm R}} a_{\rm R}} +
\sigma =0 \label{tcx}\\
\sqrt{K_{2 {\rm S}} a_{\rm S}} + \sqrt{K_{2 {\rm R}} a_{\rm R}} +
\sigma =0, \label{tcy}
\end{eqnarray}
for $ \eta_x $ and $ \eta_y $, respectively. 
Because $ K_{1 \mu } > K_{2 \mu} > 0 $  ($ \xi_{1 \mu}^2
> \xi_{2 \mu}^2 $) the nucleation will occur for the $
y$-component. To obtain an instability temperature $ T^* $ above $ T_{c {\rm
S}} $ requires that  
\begin{equation}
\xi_{2 {\rm R}}(T_{c {\rm S}}) a_{\rm R} (T_{c {\rm S}}) < d
\alpha_{\sigma} (T_{c \sigma} - T_{c  {\rm S}}).
\label{basic-cond}
\end{equation}
This simple relation gives a good insight on the basic problem
of the nucleation of local superconductivity at the
interface. Obviously, the larger $ d $ and the higher $ T_{c \sigma}
$, the  higher is $ T^* $. However, the presence of the normal-metal 
Ru tends to suppresses  
superconductivity, in particular if the coherence length $ \xi_{2 {\rm R}} $
increases. Note that if both sides of the interface were
Sr$_2$RuO$_4$, we always would find local superconductivity at a
temperature higher than $ T_{c {\rm S}} $. Furthermore,  
the suppression by the Ru-metal is weaker, if the interface were
less transparent leading to a discontinuity of the order
parameter. Reduced transparency of the interface would not change the 
dominance of the $ \eta_y $-component at this interface. The reason is 
that the order parameter component $ {\bf n} \cdot \mbox{\boldmath $
\eta $} $ is suppressed by a reflecting interface.
In any case the degeneracy of $ \eta_x $ and $ \eta_y $ of the bulk
region is lifted at the interface, since it corresponds to a region of 
effectively reduced symmetry.\cite{OGAWA}
The dominance of the $ \eta_y$ component is also plausible from a
microscopic point of view. The opening of a gap for momentum
directions along the interface leads to the gain of condensation
energy for quasiparticles with momenta parallel to the interface.
These are the quasiparticles spending the longest time in the interface
region. 
  
We assume from now on that the condition Eq.(\ref{basic-cond}) is
satisfied and the resulting transition temperature is $ T^* \approx 3
K $ corresponding to the $ 3K $-phase.  Note that the transition
temperature does not depend on the orientation of the normal vector as
long as it lies in the basal plane. The nucleating order parameter $
\mbox{\boldmath $ \eta $} $ is perpendicular to the normal vector,
i.e., $ {\bf n} \times \mbox{\boldmath $ \eta $} $. The situation does
not change much if a small $z$-axis components of the normal vector is
introduced and the properties of the local superconducting state is
determined by its inplane components.

Because the 3K-phase does not break time reversal symmetry in contrast 
to the bulk phase below $1.5$K, a further second order phase
transition has to occur at a temperature $ T^*_2 $ which is
strictly larger than the bulk $ T_c $. 
This is a different concept from conventional inhomogeneous
superconductors, where superconductivity nucleates locally on small
``islands'' which with lowering temperature increase their 
overlap and, finally, form a bulk superconducting phase without 
further symmetry breaking (apart from the percolation transition).
The temperature $ T^*_2$ depends on the coupling between the order
parameter components described by the fourth order terms in the
free energy. The two pairing components may ``attract'' or ``repel''
each other, determined by 
the sign of the parameter $ \tilde{b}= 2 b_1 - b_2 + b_3
$ ($ \tilde{b} > 0 $ repulsion and $ \tilde{b} <0 $ attraction). The
weak coupling approach leads to $ \tilde{b} > 0 $, which suppresses
the appearance of the second component. 
Hence $ T^*_2 $ may be rather close to $ T_c $ and difficult to
distinguish experimentally from the bulk transition. 

The discussion of the second transition and the form of
the order parameter in the 3K-phase is complex as it involves also
vector potential due to the presence of spontaneous currents in the
time reversal symmetry-breaking phase. Thus we turn to the numerical
solution of the complete set of GL equations including the complete set of
GL equations for the order parameter and the vector
potential. We use the parameters
$ \alpha_{\rm S} = \alpha_{\rm R} = \alpha_{\sigma}
$ and $ T_{c \sigma} = 2.8 T_{c {\rm S}} $.
Except for $ K_{5 \mu} $ we choose all coefficients 
$ K_{j\mu} $ to be the same in Ru as well as in Sr$_2$RuO$_4$:
$ K_{1 \mu}/ \alpha_{\mu} T_{c {\rm S}} = 3
K_{i \mu}/\alpha_{\mu} T_{c {\rm S}}= 1 $  $(i=2,3,4) $, which
corresponds to the basic zero-temperature coherence length $ \xi_{0}
\approx 1 $. The coefficient for the $z$-axis gradient is small in
Sr$_2$RuO$_4$, $ K_{5 {\rm S}}/\alpha_{\mu} T_{c {\rm S}} = 10^{-2} $, 
while we take, for simplicity, $ K_{5 {\rm R}} = K_{2 {\rm R}} $,
since Ru is more isotropic. The
fourth order coefficient are also chosen independent of $ x $, 
$ 2 b_{1 \mu} = 3 b_{2 \mu} = -3 b_{3\mu} = 0.4 \alpha_{\mu} $. 
We choose the interface layer thickness $ d = 1 $,
larger than it probably is in 
reality, for illustrative purpose and numerical stability. With these
parameters we obtain a nucleation temperature of $ T^* \approx 2 T_{c
{\rm S}} $ and $ T^*_2 \approx 1.33 
T_{c {\rm S}} $ which is rather high due to the large value of $ d$ .
At this transition the $ \eta_x $ component becomes finite in
addition to $ \eta_y $ and has the relative phase of $ \pm \pi/2 $. 
In Fig.~2 we show the shape of the two order parameter
components in the three different temperature regimes. Here a)
represents the genuine 3K-phase with vanishing $ \eta_x $-component
(time reversal symmetry conserving phase), b) is the intermediate time 
reversal symmetry breaking state with both components finite. Finally, 
c) is the bulk superconducting phase of Sr$_2$RuO$_4$. Note, that both 
cases b) and c) possess a complex order parameter texture at the
interface which will
be important for the magnetic properties discussed in the next
section.

\begin{figure}
\begin{center}
\includegraphics*[height=5.0cm,width=6cm]{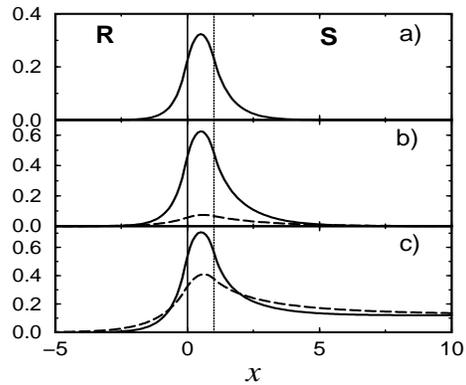}
\end{center}
\caption{Spatial dependence for the two order parameter components ($ 
|\eta_x| $: dashed line and $ |\eta_y| $ solid line): a) 3 K -phase at $ T 
= 1.75 T_{c {\rm S}} > T^*_2 $; b) intermediate time reversal symmetry
breaking phase at $ T = 1.25 T_{c {\rm S}} < T^*_2 $ 
with both components which have a relative phase $ \pm
\pi/2 $ ($ \to $ two-fold degeneracy); c) bulk superconducting phase $ 
{\bf d} = \hat{{\bf z}} (k_x \pm i k_y) $ and a texture at the interface 
at $ T = 0.95 T_{c {\rm S}} $.}
\end{figure}

\subsection{Capillary effect}

The interface instability for the superconducting state bears some 
resemblance with the wetting of a surface.\cite{WETT} Similar to
the wetting phenomena of liquids we find here capillary effects.
Modulations of interfaces on length scales comparable to the coherence 
length can enhance the nucleation temperature. While this capillary
effect is rather simple for conventional superconductors, there are
complications in the case of an unconventional superconductor. In
particular, there can be frustration effects due to the internal
structure of the pair wave function, as we will see in the next
section. Here we would like to consider first the rather simple
situation of an interface that is not flat but has a weak modulation.

The enhancement of the transition temperature on a spatial modulated 
interface can be most simply interpreted by the effective increase of surface
area. This corresponds also to an enhanced mutual overlap of the order
parameters nucleating at different points of the interface or on different
inclusions.  
Let us assume that the interface is only slightly modulated, described by
$ x_0(y) = l \sin (2 \pi y/ L') $ where $ d \ll l \ll L' \sim 
\xi_{i \mu} $.  
Then, by the most simple variational approximation we replace the interface 
term by
\begin{equation}
{\cal F}_i = \int d^3 r | \mbox{\boldmath $ \eta $}|^2 \sigma 
\sqrt{1+ \left( \frac{2 \pi l}{L'} \cos \left( \frac{2 \pi y}{L'} \right) 
\right)^2} 
\delta(x-x_0(y)).
\end{equation}
The important modification appears via the new interface metric, which 
accounts for the fact that the interface is wider or ``denser''. This leads to
an enhanced instability temperature, even if we approximate the spatial
dependence of the order parameter by a form like in
Eq.(\ref{exps},\ref{expr}). 
Using this variation approach the instability equation for $ \eta_y $ 
changes to
\begin{equation}
\sqrt{K_{2{\rm s}} a_{{\rm S}}} 
+\sqrt{K_{2 {\rm R}} a_{{\rm R}}} + \sigma R = 0,
\end{equation}
where the factor $ R$ is
\begin{equation}
R = \frac{1}{L'}\int_0^{L'} d y \sqrt{1+ \left( \frac{2 \pi l}{L'} \cos 
\left( \frac{2 \pi y}{L'} \right) 
\right)^2} \approx 1 +  \left( \frac{\pi l}{L'}\right)^2 .
\end{equation}
Obviously, $ R $ is always larger than 1 
and leads to an effective enhancement of
$ \sigma $. We have ignored in our variational approach the spatial
dependence of the order parameter along the interface. 
Including this aspect lead to even further enhancement of the
nucleation temperature. 

Another aspect of the capillary effect is the mutual influence of
interfaces which we would like to consider on the example of
two parallel interfaces. To be specific we assume a thin Ru-metal slab
of thickness $ L $ sandwiched between Sr$_2$RuO$_4$, again with normal 
vector parallel to the $ x$-axis. Both interfaces
have the same properties. Then it is easy to derive the instability
equations taking the boundary conditions into account. 
There are two combinations of the nucleating
order parameters on the two interfaces: a ``bonding'' and 
``antibonding'' configuration which is even or odd, respectively,
under reflection at the center of the slab. 
For the instability equations of the
$ \eta_y $-component we obtain,
\begin{equation}
\sqrt{K_{2{\rm s}} a_{{\rm S}}} 
+ {\rm tanh}\left(\frac{L}{2 \xi_{2 {\rm R}}} \right) \sqrt{K_{2 {\rm
R}} a_{{\rm R}}} + \sigma = 0 
\end{equation}
for the bonding and
\begin{equation}
\sqrt{K_{2{\rm s}} a_{{\rm S}}} 
+ {\rm coth}\left(\frac{L}{2 \xi_{2 {\rm R}}} \right) \sqrt{K_{2 {\rm
R}} a_{{\rm R}}} + \sigma = 0 
\end{equation}
for the antibonding combination. 
Obviously, we recover in both cases 
the original instability equation Eq.(\ref{tcy}), if we
separate the two interfaces far apart, $ L \gg \xi_{2 {\rm R}} $. 
If, however, $ L \sim \xi_{2 {\rm R}} $ then the second (positive) term is
diminished (enhanced) and a higher (lower) transition temperature $
T^* $ results in the case of bonding (antibonding). Naturally, the
same kind of capillary effect occurs also in the inverse situation
where a Sr$_2$RuO$_4$ slab is surrounded by Ru. 
The bonding combination of the two interface states just corresponds
to the adjustment of the phases of two superconducting islands. The
antibonding combination is equivalent to a phase difference of $ \pi
$ and is the energetically least favorable case. We will see in the
next section that this aspect is important for 
frustration effects of the 3K phase.  

We can conclude that the onset of the 3K phase is rather inhomogeneous 
on the interfaces as well, because the capillary effects have a strong
influence on the nucleation of the superconducting order parameter. 

\section{Physical consequences}

\subsection{Spontaneous interface magnetism}

While the onset of superconductivity is visible in the reduction of
electrical resistance, the significant features of second transition
at $ T^*_2$ are less easy to detect. Bulk superconducting
double transitions are often observed through 
specific heat anomalies. Here the only significant anomaly,
however, is seen at the onset of bulk superconductivity at $ T_c = 1.5 K$. 
Magnetic properties may, on the other hand, allow us to observe the
onset of the superconducting phase at $ T^*_2 $
that violates time reversal symmetry.
The solution of the full GL equations
show that the below $ T^*_2 $ spontaneous supercurrents occur parallel 
to the interface. In Fig.~3 we show the magnetic field and current
distribution at the interface for the case corresponding to Fig.~2b. 
There are two currents along the interface flowing in opposite
direction as a result of the texture of the two order parameter
components at the interface. Screening currents are weak due to absence of
superconductivity in the bulk. Thus, the field resulting from the
currents is strongly peaked on the interface and can even generate an
overall finite magnetic flux. Since the flux is finite, it could be 
observed, in principle, by high-resolution magnetic microscopes, such as
scanning Hall probes or a SQUID microscopes. Another sensitive probe for
local magnetic field distributions is zero-field muon spin relaxation,
which is expected to show an increase of the depolarization rate below
$ T^*_2 $. However, if $ T^*_2$ is very close to $ T_c $ (as is very
likely the case) then the signal will be obscured by the bulk
superconducting state which has previously been observed by $ \mu
$SR.\cite{LUKE} The muon spin relaxation experiment would in any case
give a distinction between magnetic properties of the 3K-phase at $
T^* $ and the bulk phase. 

\begin{figure}
\begin{center}
\includegraphics*[height=5cm,width=6.5cm]{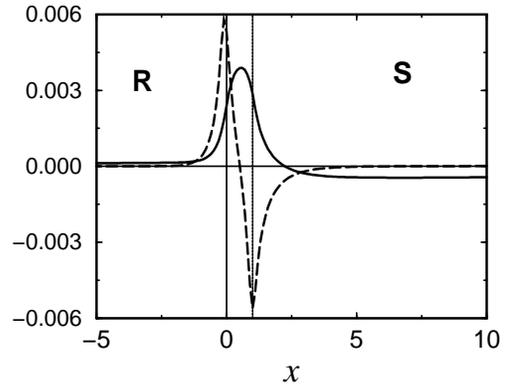}
\end{center}
\caption{Spontaneous magnetic field $ B_z $ (solid line) and
supercurrent $ j_y $ (dashed line) distribution for the state given in
Fig.2b) at $ T = 1.25 T_{c {\rm S}} $.} 
\end{figure}

\subsection{Frustration effects in the inhomogeneous state}

We have seen in the previous section on the capillary effect 
that the lowest energy configuration of order parameters nucleated 
on different interfaces naturally corresponds to equal order parameter 
phases. Let us now
see how the superconducting state would arrange on several inclusions
close enough to each other that their order parameters substantially
overlap. If the order parameter were conventional, it would be easy to
adjust all the order parameter phases on the different 
inclusion to be equal. This is not the case, if the order parameter
has $ p $-wave symmetry as is illustrated in Fig.~4. The three
Ru-inclusions depicted in Fig.~4 carry a superconducting state 
which on every point of the interface correspond to a $ p $-wave state 
with the momentum direction aligned with the interface, i.e. the gap
node lies always parallel to the normal vector. The interface regions on
different inclusions close to
each other behave like junctions or weak links connecting a network of 
the superconducting islands.

The direction of the positive lobe of the 
pair wave function is indicated by the
arrows in Fig.~4 where we assume that the order parameter 
phase is constant on
each inclusion. Parallel (antiparallel) arrows on neighboring
interfaces correspond to bonding (antibonding) configurations of the
overlapping order parameters, or
equal phase ($ \pi $-shifted phases). It is obvious that in the
arrangement of the order
parameter on each inclusion introduces a $ \pi $-phase shift between 
its weak links. Although we may change the phases on each island, 
it is impossible to adjust them so that all links have a zero phase
difference for loops consisting of an odd number of inclusions (Fig.~4
shows this for the case of three inclusions). The situation is in many
respects similar to the case described for granular $d$-wave
superconductors.\cite{SVED,BRAUNISCH,PME} We may identify loops which
contain $ \pi $ phase shifts and are 
consequently frustrated, since they cannot adjust the order parameter
phase to minimize the energy of all weak links simultaneously. 
The frustration can be released by introducing phase gradients
(supercurrents) which are energetically favorable, if the weak links
grow strong enough. This would yield orbital magnetic moments. This
effect results from the competition between weak link energy and 
magnetic field energy (determined by geometry) 
as in usual frustrated $ \pi $-loops. However, the frustration
also is reduced with the time reversal symmetry breaking transition
at $ T= T^*_2 $. Both ways have similar impact by diminishing
frustration and lead to spontaneous currents.
In the network of superconducting islands the transition to the time
reversal symmetry breaking state is inhomogeneous. 

\begin{figure}
\begin{center}
\includegraphics*[height=5.0cm,width=6cm]{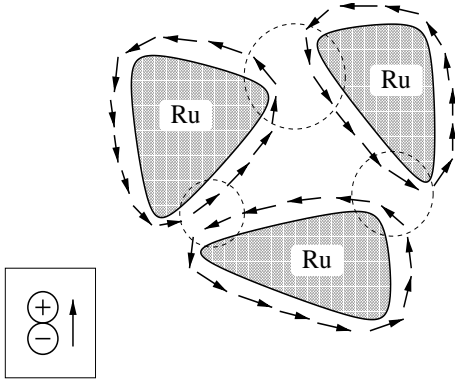}
\end{center}
\caption{Configuration of the $ p $-wave pair wave function on
Ru-inclusions (shadowed regions) in a crossection parallel to the
$x$-$y$-plane. The arrow indicates the direction of
positive wave function (see box). The dashed circles show the regions 
forming weak links between inclusions. The fact that the arrows of
different inclusions in the weak link region are antiparallel
corresponds to a phase difference of $ \pi $. In the given
configuration there is no way to adjust the phases of all three
inclusions so that all weak links have minimal energy, i.e. a
vanishing phase difference. Therefore this system is frustrated. 
}
\end{figure}

The frustrated 3K-phase with spontaneous currents can exhibit
enhanced absorption in the ac-susceptibility. Dissipative processes,
such as phase slips,  are associated with the hysteretic reversal of the
spontaneous currents. Thus the absorption would 
be sensitive to the application of a small static external fields
which would bias the spontaneous currents (a similar  situation
was observed in granular high-temperature 
superconductors\cite{SVED,BRAUNISCH,PME}) The presence of spontaneous
orbital magnetic moments in the inhomogeneous time reversal
symmetry violating phase above $ T_{cS} $ can cause a characteristic
non-linear magnetic response in small external fields, 
similar to the paramagnetic Meissner effect
or Wohlleben effect in high-temperature
superconductors.\cite{SVED,BRAUNISCH,PME} Under field-cooling conditions
the magnetic moments would be generated in a polarized way yielding
a paramagnetic contribution. Probably, 
the signal would be considerably weaker than in high-temperature
superconductors. The inhomogeneous 3K-phase is, however, more suitable
to measure the spontaneous magnetism than the bulk superconducting
phase, where the macroscopically visible magnetism is only resulting
from surface currents, while it is screened or compensated 
in the interior of the systems. 

\subsection{Quasiparticle spectrum}

The presence of superconductivity modifies the quasiparticle spectrum in
small enclosed normal-metal regions through the formation of
so-called Andreev bound states. Andreev bound states are standing
waves of an electron-hole superposition, e.g. in a normal
metal region enclosed by a superconductor. In a quasiclassical picture
the standing wave corresponds to an electron and hole travel on the
same classical (ballistic) 
trajectory, but in opposite direction and are subject
to Andreev reflection upon impact in the superconductor at both ends
of the trajectory. The energy of such a standing wave depends on the
phase of the gap functions at the boundaries of the classical
trajectories and can be estimated by Bohr-Sommerfeld quantization.
Energies much smaller than the gap of the superconductor are
approximately given by,
\begin{equation}
E(n,\chi) \approx \frac{v_F}{2 L} [ (2 n +1) \pi + \chi]
\label{Bohr-Sommerfeld}
\end{equation}
where $ v_F $ is the Fermi velocity, $ L $ the length of the
trajectory, $ n$ an integer and, most important, $ \chi $ is the 
difference between the phases of the gap functions at both ends of the 
ballistic trajectory (for simplicity
we ignore here the effect of impurity scattering
which can also be the origin low-energy states). 
For unconventional superconductors the phase
difference $ \chi $ depends also on the direction of the electron (hole)
momentum, because the gap function, in general, has an anisotropic
phase structure, i.e. a different phase for different directions of
momenta on the Fermi surface. Thus, even in case of a homogeneous
order parameter, the phase $ \chi $ appearing in the Andreev
scattering process would no trivial, but revealing the internal phase  
structure of the pairing state. For real order parameters we find
$ \chi = \pm \pi $ which is most important, because it leads to a
zero-energy state independent of $ v_F $ and $ L $. This phase
difference appears, if the gap function has positive and negative
sign depending on the momentum direction on the Fermi surface. 
From the quasiclassical point of view underlying
Eq.(\ref{Bohr-Sommerfeld}) the geometry of the classical trajectory
decides about the phase $ \chi = 0 $ or 
$\pi $. Since a large fraction of trajectories can have $ \chi = \pi $ 
and all of them yield a zero-energy state, 
the zero-energy states ($ \chi = \pi $) lead to an enhanced
density of quasiparticle states at the Fermi surface.
The presence of
this kind of zero-energy bound state has been intensively investigated
for surfaces of $ d $-wave supercondcutors.\cite{HU,TANAKA} 

In our case a $ p $-wave pairing state nucleated at the interface
generates a gap function which has regions of different sign on the Fermi
surface. Zero-energy Andreev bound states are likely to
occur within the Ru-inclusions as well as in Sr$_2$RuO$_4$ between 
Ru-inclusions that are sufficiently close to each other. We encounter
here a complex form of an (imperfect) Andreev billiard. The
Andreev reflections are not perfect. Due the finite width of the
superconducting regions, the zero-energy level acquires some
width. Nevertheless, there is a strongly enhanced density of states at
zero energy. 
Once time reversal symmetry is broken the phase differences $ \chi $ for 
trajectories deviate from 0 and $ \pi $. The originally enhanced 
density of states at zero-energy spreads over a larger energy region. 
(Note that the reduction of the density of states at the Fermi energy
represents also a driving force for the time reversal symmetry
breaking transition, as was suggested also for the surface states of a 
$ d $-wave superconductor.\cite{MATSUSHIBA,FOGEL}) 

One experimental indication for the enhanced density of states 
zero-energy states is the observation of so-called zero-bias
anomaly, i.e. an increased quasiparticle tunneling conductance at zero
voltage. Recent measurements of quasiparticle tunneling conductance in
c-axis facing break junctions of Sr$_2$RuO$_4$ with Ru-inclusions
report the observation of zero-bias anomalies.\cite{MAO} In the
3K-phase a zero-bias anomaly in current-voltage characteristics
develops gradually with decreasing temperature. This conductance peak 
deforms into a pronounced bell shape combined with a residual zero-bias
anomaly, below the bulk superconducting transition. Assuming that these
tunneling features reflect the quasiparticle density of states
connected with the Andreev billiard, we can interpret them within the
scenario of the nucleated $ p $-wave state which turns into a
time reversal symmetry breaking state at lower temperature. We would
like to emphasize the fact that break junctions in samples without
Ru-inclusions did not show any similar features above and below the
onset of bulk superconductivity at 1.5 K. Note also that
an analogous phenomenon, the deformation of the zero-bias anomaly, has been
observed for tunneling conductance into the [110]-surface of
YBa$_2$Cu$_3$O$_7$.\cite{COVING} Also in that case broken time
reversal symmetry is most likely cause.\cite{MATSUSHIBA,FOGEL}

It is clearly desirable to have more extensive experimental
investigation of the quasiparticle spectrum in and around the
Ru-inclusions via tunneling and related
probes.\cite{YAMA,HONER,MATSUSIG} In particular, 
using a contact to a single inclusion and measuring the quasiparticle
current-voltage characteristics through the Ru-metal inclusion into 
Sr$_2$RuO$_4$ could provide further valuable information on the 3K as
well as bulk superconducting phase. A detailed analysis of the
phenomena discussed in this section will be given elsewhere. 

\section{Upper critical field}

The 3K phase has an upper critical field which
exceeds the bulk critical field considerably and shows a weaker
anisotropy of in-plane versus out-of-plane critical field.
The fact that superconductivity is confined in a small layer at the
interface naturally leads to an reduction of orbital depairing, if the
external magnetic field is applied parallel to the interface. 
This type of effect is known for superconductivity at 
other planar defects such as twin boundaries. \cite{BULA,ABRIKOSOV}  
In a magnetic field superconductivity nucleates first at interfaces
which are parallel to the field. We will analyze this effect
for the [100]-interface with magnetic fields $ {\bf B} \perp x $-axis.
Unfortunately, a complete analytical treatment of this problem is not
possible even for the GL formulation. Therefore we will
restrict our analytic discussion to the region of very small fields at the
onset of the 3K-phase and use a
variational approach. This allows us to illustrate a few basic
features of the upper critical field. Then we will consider the
behavior of $ H_{c2} $ by numerical means for the model introduced
above. 

For temperatures close to $ T^* $ the critical field is small and we
may consider it as a perturbation.\cite{ABRIKOSOV} Thus, we will use
the exponential form of the order para\-meter appearing in zero-field
at the onset of the 3K-phase: $ \eta_{x \mu} = \eta_{x0} \exp(-|x|/
\xi_{1 \mu}) $ and $ \eta_{y \mu} = \eta_{y0} \exp(-|x|/ \xi_{2 \mu})
$ with $ \xi_{i \mu} = K_{i \mu}/a_{\mu} $. The magnetic field lies in
the plane of the interface $ {\bf B} = H (0, \sin \theta, \cos \theta
) $ with the vector potential $ {\bf A} = H (x-x_0) ( 0, - \cos
\theta, \sin \theta ) $.  While in zero-field only the $ \eta_y
$-component is nucleated at $ T^* $, in general both components can
appear in a finite magnetic field. We use now the given form of the
order parameter and calculate the free energy in a finite field by
integrating over the spatial coordinates. In this first-order
perturbative form the free energy per unit area of the interface is
then up to second order in the the order parameter,
\begin{equation} \begin{array}{ll}
{\cal F}_{\rm var} = & \displaystyle \sum_{\mu={\rm S},{\rm R}} \left
[ |\eta_{x0}|^2 \left\{ a_\mu \xi_{1 \mu} + \frac{\sigma}{2} + 
\frac{\gamma^2 H^2}{4} \xi_{1 \mu}^3 f_{x \mu}(\theta) \right\} \right. \\ &
 \displaystyle + |\eta_{y0}|^2 \left\{ a_\mu \xi_{2 \mu} +
+ \frac{\sigma}{2} + \frac{\gamma^2 H^2}{4} \xi_{2 \mu}^3 
f_{y\mu}(\theta) \right\}  \\ & 
\displaystyle \left.    + i(\eta_{x0}^* \eta_{y0} - \eta_{x0} \eta_{y0}^*)
r \gamma H  K_{2\mu} \xi_{2 \mu}^3 \cos \theta  \right]
\label{fevar}
\end{array} 
\end{equation}
with $ \gamma = 2e/\hbar c $ and $ r = (\sqrt{3}-1)/(\sqrt{3}+1)^2 $
where we used the relation $ \xi_{1 \mu} = \sqrt{3} \xi_{2 \mu} 
$, keeping the weak-coupling relations among different $ K_{i\mu} $
for cylindrical symmetry. To simplify the
free energy (\ref{fevar}) we have fixed $ x_0 $ to zero, since it
plays a minor role for our discussion. The 
anisotropy parameters are  
\begin{equation} \begin{array}{l}
f_{x \mu} (\theta) = K_{2 \mu} \cos^2 \theta + K_{5 \mu} \sin^2 \theta ,
\\ 
f_{y \mu} (\theta) = K_{1 \mu} \cos^2 \theta + K_{5 \mu} \sin^2 \theta .
\end{array}
\end{equation}

First we consider the case of the in-plane field ($ \theta = \pi/2
$). Here only the $ \eta_y $-component appears, which corresponds to the
polar state with its nodes perpendicular to the 
field, which also appears in the bulk superconducting phase for
in-plane fields.  The zero of the corresponding coefficient in $ {\cal
F}_v $ defines the instability, which leads to
\begin{equation}
H_{c2} = \frac{\Phi_0}{\pi} \left[ - d\frac{2 a^*_y(T)}{K_{5R}
\xi_{2R}^3} \right]^{\frac{1}{2}} 
\end{equation}
where we used that $ K_{5S} \ll K_{i \mu}, K_{5 R} $ ($ i \neq 5$). In
this case including a finite value of $ x_0 $ we find that $ x_0 = -
\xi_{2R}/2 $ maximizes $ H_{c2} $ 
(we omit here an explicit demonstration). Note that we can write $
\sum_{\mu} ( \xi_{2 
\mu} a_{\mu}(T) + \sigma (T)/2) = a^*_y (T) $ which is proportional to 
$ (T-T^*) $ close to $ T^* $. Consequently, we observe a square-root
dependence $ H_{c2} \propto |T^* - T|^{\frac{1}{2}}
$. \cite{BULA,ABRIKOSOV}  For curved interfaces the 
regions tangential to the field tend to allow the nucleation at higher 
temperature. Simultaneously, we have, however, to include capillary
effects. Thus, the observation of a pure square-root temperature
dependence would be masked in reality. 

\begin{figure}
\begin{center}
\includegraphics*[height=5cm,width=6.5cm]{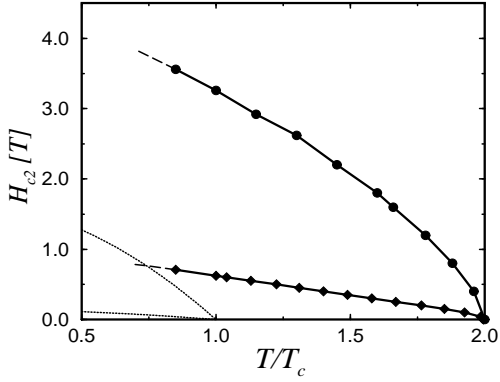}
\end{center}
\caption{Upper critical fields for the 3K and bulk phase: 
The numerical solution of the GL equations give the upper 
critical fields of the 3K-phase: inplane (circles) and $z$-axis
(diamonds). The corresponding bulk critical fields are given by doted
lines. Note, the weaker anisotropy of the critical field of the
3K-phase. The numerical data compare well with the experimental data
on a qualitative as well as quantitative level.\cite{3K,ANDO}} 
\end{figure}

For the field parallel to the $z$-axis ($ \theta =0 $) the $
\eta_x $-component is involved too, despite the lower critical
temperature in zero field. The optimal value of $ x_0 $ is very close
to zero so that we fix 
$ x_0 = 0 $. With this simplification the instability equations
involving the coupling of both components have the form, 
\begin{equation} \begin{array}{l} \displaystyle 
\{a^*_x(T) +  \left( \frac{\gamma H}{2} \right)^2 \Lambda_x
\} \eta_{x0} - i \gamma H \Lambda'\eta_{y0} =0 \\   \displaystyle 
i \gamma H \Lambda' \eta_{x0} + \{a^*_y(T) + \left( \frac{\gamma H}{2}
\right)^2 \Lambda_y  \} \eta_{y0} =0 \\
\label{zaxis-instab}
\end{array}
\end{equation}
where $ \Lambda_x = \sum_{\mu} K_{2 \mu} \xi^3_{1 \mu} $, 
$ \Lambda_y = \sum_{\mu} K_{1 \mu} \xi^3_{2 \mu} $ and $ \Lambda'= r
\sum_{\mu} K_{2 \mu} \xi_{2 \mu} $. Moreover, $ a^*_x (T) = 
\sum_{\mu} ( \xi_{1 \mu} a_{\mu}(T) + \sigma (T)/2) $ is the effective
second order coefficient for the $ \eta_x $-component and changes sign 
at a temperature $ T_{cx} $ which lies between $ T^*_2 $ and $ T^* $.
The critical field is obtained by searching the non-trivial solution of
this equation system, 
\begin{equation}
H_{c2}(T) \approx \frac{\Phi_0}{\pi} \left[ - \frac{ a^*_y(T)}{\Lambda_y}
\frac{ a^*_x(T)}{a_x^*(T)- 4 \Lambda'^2 /\Lambda_y}
\right]^{\frac{1}{2}}
\label{hc2-caxis}
\end{equation}
where we used $ |a^*_y (T)| \ll a^*_x(T) $ for $ T \to T^* $. Thus we
find again a basic square root behavior of $ H_{c2} (T) $ as in the
case of inplane fields. However, in addition we see that the coupling
of the two components yields an enhancement factor to $ H_{c2} $ which
becomes stronger as the temperature is lowered. (Note, however, that the
present form is only valid for $ 0 < a^*_x(T) \gg 4
\Lambda'^2/\Lambda_y $ and in any case our variational approach is
applicable only in a restricted region close to $ T^* $.)
This enhancement is important, since it may modifies the overall form
of the temperature dependence of $ H_{c2} $. Before considering this point
numerically let us, however, address the issue of $ H_{c2} $
anisotropy. 

For the bulk superconducting phase of Sr$_2$RuO$_4$ the critical
fields are given by
\begin{equation}
H_{c2}^{\perp} = \frac{- \Phi_0 a_{\rm S}(T) }{4 \pi K_{2 {\rm S}}}
\mbox{ and } 
H_{c2}^{\parallel} = \frac{- \Phi_0 a_{\rm S}(T)}{2 \pi \sqrt{K_{2
{\rm S}} K_{5 {\rm  S}}}}
\end{equation}
with $ T < T_{c {\rm S}} $ and ignoring anisotropy of $ H_{c2} $ for
different inplane orientations, which is small.\cite{REV} This
leads to an anisotropy factor 
of order $ H_{c2}^{\parallel}   
/ H_{c2}^{\perp} \approx  2 \sqrt{K_{2 {\rm S}}}/\sqrt{K_{5 {\rm S}}} $
whose experimental value is about 12 for $ T $ close to $ T_{c {\rm S}} $ in
experiments ($ K_{2{\rm S}} \gg K_{5{\rm S}} \approx 0.03 K_{2{\rm S}} 
$).\cite{3K}  
On the other hand, the anisotropy for the 3K-phase may be 
less anisotropic with a factor $ 4 \sim 5 $,\cite{3K,ANDO} since 
\begin{equation}
\frac{H_{c2}^{\parallel}}{H_{c2}^{\perp}} \approx \left(\frac{2(\xi_{2S}^3
K_{1S} + \xi^3_{2R} K_{1R})}{\xi_{2R}^3 K_{5R}}
\right)^{\frac{1}{2}}
\label{ratio2} 
\end{equation}
gives a smaller ratio due to the fact that Ru-metal is basically
isotropic with $ K_{5 {\rm R}} \sim K_{2 \mu} $ and also the
coherence lengths at $ T=T^* $ are of the same order in Ru and
Sr$_2$RuO$_4$. The reduction of anisotropy originates from
the isotropy of 
the Ru-metal which leads to a stronger coupling of the order parameter to 
the inplane field than for Sr$_2$RuO$_4$. 

The numerical evaluation of the upper critical field for the model
used in Sect.2.2 illustrates the temperature dependence on qualitative 
level. The result are shown in Fig.~4 
for the assumption that $ K_{5 {\rm S}} = 0.03 K_{2 {\rm S}} $ and
$ K_{5 {\rm R}} = K_{2 {\rm S}} $. The circles and diamonds are the
numerical results for the critical field inplane and along the
$z$-axis, respectively. We observe indeed a weaker
anisotropy for the critical field of the 3K-phase than for the bulk
phase also indicate in Fig.~4. The initial temperature dependence has
square root dependence as obtained analytically. While the inplane critical
field has a downward curvature in the plotted temperature range,
the $z$-axis field is nearly linear. The analytical
expression in Eq.(\ref{hc2-caxis}) indicates that even the upward
curvature would be possible for the $ z $-axis critical field. 
Our result is close to the experimental data
qualitatively and to some extent quantitatively, with the choice of
parameters giving $ T^* \approx 2 T_{c} $ and the proper anisotropy 
of the bulk critical fields.\cite{3K,ANDO} 
The peculiar temperature dependence of the $z$-axis
critical field of the 3K-phase is also similar to that found in
experiment.\cite{3K,ANDO} 

\section{The $s$-wave order parameter}

Ru metal becomes a conventional superconductor at a transition
temperature of about 0.5 K. Thus, we may include an s-wave order
parameter into our theory. The coupling between the spin-singlet 
order parameter $ \eta_s $ and the spin-triplet order parameter 
$ \mbox{\boldmath $ \eta $} $ requires that we 
take spin-orbit coupling
into account. The difference between spin-orbit coupling
in Ru and Sr$_2$RuO$_4$ yields spin flip tunneling at the interface
which yields a coupling between $ \eta_s $ and $ \mbox{\boldmath $
\eta $} $.\cite{GESH,REV} The coupling term has the form
\begin{equation}
{\cal F}_{s.o.} = \int_{\rm interface} dS [\eta_s^* ({\bf n} \times 
\mbox{\boldmath $ \eta $} + c.c.] 
\end{equation}
where $ \eta_s $ is the $s$-wave order parameter on the Ru-side and 
$ \mbox{\boldmath $ \eta $} $ the two-component $p$-wave order parameter on
Sr$_2$RuO$_4$-side, $ {\bf n} $ is the interface normal vector and $ t 
$ is a coupling constant. Obviously, for $ {\bf n} = (100) $ only the
$ \eta_y $-component couples. 

Further there is a coupling between the two order parameters also
away from the interface which is again due to spin-orbit coupling. 
The corresponding term in the free energy, derived based on symmetry
arguments, has the form,
\begin{equation}
{\cal F}_{sp} = \int dV \tilde{a} [ (D_x \eta_y - D_y \eta_x)^* \eta_s 
+ c.c.] 
\end{equation}
where $ \tilde{a} $ is again a coupling constant different on the two
sides. This term is only active, if there is a spatial variation of
the order parameter or a magnetic field is present. We find that
the spatial variation of the order parameter along the $x$-axis would
couple exclusively the $ \eta_y $-component to $ \eta_s $. 

This structure of order parameter coupling leads to a support for the
nucleation of the $ \eta_y $-component on the discussed 
[100]-interface due to the mixing with the $ s $-wave order parameter
intrinsic to Ru. On a qualitative level, however, the
inclusion of an $ s $-wave component would not modify the properties
of the 3K-phase on an essential way. 
 
\section{Conclusion}

In this article we have interpreted the 3K-phase of Sr$_2$RuO$_4$ with
Ru-metal inclusion as an inhomogeneous superconducting state located
at the interface between the material phases. 
This leads to a superconducting phase is different
qualitatively from the bulk phase appearing below 1.5 K. The 3K-phase
conserves time reversal symmetry. This implies an additional phase
transition where time reversal symmetry is spontaneously 
broken. This second transition occurs above the onset of bulk
superconductivity. Irregular shapes and distribution of 
the Ru inclusions and the capillary effects would prevent a very 
sharp transition for the onset of the 3K-phase as well as the second
transition. 

The origin of multiple superconducting transitions is analogous to 
that of the splitting of the superconducting phase transition of
degenerate order parameters by applying symmetry-lowering uniaxial
stress.\cite{STRESS,OGAWA} Indeed the interface represents a region of the
system where the symmetry is effectively lower, lifting the degeneracy 
between the two order parameters $ \eta_x $ and $ \eta_y $. Therefore
the experimental proof of the conservation of time reversal symmetry in the
3K-phase would be a clear confirmation of a having a two-component
order parameter.

We have argued that the phase structure of the pair wave function
can lead to frustration effects in the coupling of the
superconducting order parameters on different Ru-inclusions or
different regions on the same inclusion. This is close related with
the fact that this type of order parameter can lead to low- or zero-energy 
Andreev bound states, since the configuration of inclusions forms a 
complex Andreev billiard system. 
There are various experimental consequences due to these properties,
some of which have been partially already investigated experimentally.
(1) The Ru-metal inclusions provide an interesting
way to tunnel into Sr$_2$RuO$_4$. Contacts via Ru inclusions
could reveal more about the structure of the quasiparticle
spectrum.\cite{MAO} (2) The superconducting interface states of
different neighboring Ru-inclusions overlap and form a
complex network. The study of the critical current as a function of
temperature may give another tool to investigate the unconventional
nature of this state.   
In particular, the frustration effects mentioned above and the
spontaneous currents which occur in the time reversal symmetry breaking 
phase can yield characteristic anomalous behavior.
(3) The low-field magnetic response may show strong non-linear behavior
and cooling history-dependence in the time reversal symmetry breaking state
for temperatures above $ T_{c {\rm S}} $, a phenomena which could be
rather similar to the paramagnetic Meissner (or Wohlleben) effect
in granular high-temperature superconductors.\cite{SVED,BRAUNISCH,PME} 
(4) The study of microwave absorption in a small static magnetic field 
may show a non-monotonic field dependence as in some granular
high-temperature superconductors.\cite{SVED,BRAUNISCH,PME}

These is only a selection of possible unusual properties of the
3K-phase. It is obvious that this phase and its properties provide a
very good tool to investigate the superconductivity in Sr$_2$RuO$_4$ 
from a new point of view. 

We would like to thank Y. Maeno, Z.Q. Mao, T. Akima, M. Wada, H. Yaguchi, 
Y. Liu and A. Furusaki 
for many stimulating discussions concerning the 3K-phase. 
This work was supported by a Grant-in-Aid of the Japanese Ministry of
Education, Science, Culture and Sports. H.M. is also grateful for the
support by a long-term fellowship of Japan Society for the Promotion
of Science.


\begin{thebibliography}{99}
\bibitem{MAENO1}
Y. Maeno et al.: Nature {\bf 372} (1994) 532.
Physica C {\bf 282-287} (1997) 206.
\bibitem{NATRICE}
T.M. Rice: Nature {\bf 396} (1998) 627. 

\bibitem{RICE} T. M. Rice and M. Sigrist:
J. Phys. Condens. Matter. {\bf 7} (1995) L643; 
G. Baskaran, Physica {\bf B 223-224} (1996)  490.

\bibitem{ISHIDA1} K. Ishida et al.: Nature {\bf 396} (1998) 658.

\bibitem{LUKE} G.M. Luke et al.: Nature {\bf 394} (1998) 558.

\bibitem{3K} Y. Maeno et al., Phys. Rev. Lett. {\bf 81} (1998) 3765.

\bibitem{ANDO} T. Ando, T. Akima, Y. Mori and Y. Maeno,
J. Phys. Soc. Jpn. {\bf 68} (1999) 1651.

\bibitem{NEUTRON} M. Braden, W. Reichardt, S. Nishizaki, Y. Mori and
Y. Maeno, Phys. Rev. {\bf B57} (1998) 1236. 

\bibitem{LDA} T. Oguchi, Phys. Rev. {\bf B51} (1995) 1385;
D.J. Singh, Phys. Rev. {\bf B52} (1995) 1358. 

\bibitem{TERAKURA} R. Matzdorf, Z. Fang, J. Zhang, T. Kimura,
Y. Tokura, K. Terakura and E.W. Plummer, Science {\bf 289} (2000)
746.

\bibitem{fluxdist} P.G. Kealey et al., Phys. Rev. Lett. {\bf 84}
(2000) 6094. 

\bibitem{OGAWA} M. Sigrist, N. Ogawa and K. Ueda,
J. Phys. Soc. Jpn. {\bf 60} (1991) 2341.

\bibitem{WETT} E. Montevecchi and J.O. Indeken, cond-mat/0009328.

\bibitem{HU} C.R. Hu, Phys. Rev. Lett. {\bf 72} (1994) 1526.

\bibitem{TANAKA} Y. Tanaka and S. Kashiwaya, Phys. Rev. Lett. {\bf
74} (1995) 3451.

\bibitem{MATSUSHIBA} M. Matsumoto and H. Shiba,
J. Phys. Soc. Jpn. {\bf 64} (1995) 1703; {\it ibid.} 3384; {\it
ibid.} {\bf 65} (1996) 2194.

\bibitem{FOGEL} M. Fogelstr\"om, D. Rainer and J.A. Sauls,
Phys. Rev. Lett. {\bf 79} (1997) 281.

\bibitem{MAO} Z.Q. Mao, K.D. Nelson, R. Jin, Y. Liu and Y. Maeno,
cond-mat/0101410.

\bibitem{COVING} M. Covington, M. Aprili, E. Paraoanu and L.H. Greene, 
Phys. Rev. Lett. {\bf 79} (1997) 277.

\bibitem{YAMA}  M. Yamashiro, Y. Tanaka and S. Kashiwaya,
Phys. Rev. B {\bf 56} (1997) 7847; J. Phys. Soc. Jpn. {\bf 67}
(1998) 3224. 

\bibitem{HONER} C. Honerkamp and M. Sigrist, J. Low Temp. Phys. {\bf
111} (1998) 895.

\bibitem{MATSUSIG} M. Matsumoto and M. Sigrist,
J. Phys. Soc. Jpn. {\bf 68} (1999) 994. 

\bibitem{BULA} A.I. Buzdin and L.N. Bulaevskii, JETP Lett. {\bf 34}
(1981) 112; A.F. Andreev, JETP Lett. {\bf 46} (1987) 584. 

\bibitem{ABRIKOSOV} A.A. Abrikosov, {\it Fundamentals of the Theory of 
Metals}, North-Holland, Amsterdam (1988).

\bibitem{GESH} V.B. Geshkenbein and A.I. Larkin, JETP Lett. {\bf 43}
(1986) 395. 

\bibitem{REV} M. Sigrist and K. Ueda, Rev. Mod. Phys. {\bf 63} (1991) 
239. 

\bibitem{SVED} P. Svedlindh, K. Niskanen, P. Nordling, P. Nordblad,
L. Lundgren, B. L\"onnberg and T. Lundstr\"om, Physica {\bf 162-164}
(1989) 1365. 

\bibitem{BRAUNISCH} W. Braunisch, N. Knauf, G. Bauer, A. Kock,
A. Becker, B. Freitag, A. Gr\"utz, V. Kataev, S. Neuhausen, B. Roden,
D. Khomskii and D. Wohlleben, Phys. Rev. {\bf B48} (1993) 4030.  

\bibitem{STRESS} M. Sigrist, R. Joynt and T.M. Rice,
Europhys. Lett. {\bf 3} (1987) 629. 

\bibitem{PME} M. Sigrist and T.M. Rice, 
J.~Phys.~Soc.~Jap.~61 (1992) 4283; Rev. Mod. Phys. 67 (1995) 503.

\end{thebibliography}
\end{document}